\documentclass[a4paper,12pt]{article}
\pdfoutput=1
\usepackage{epsfig}
\usepackage{amssymb,subfigure}
\usepackage{amsfonts}
\usepackage{amsmath}
\usepackage{euscript}
\usepackage{verbatim}
\usepackage{latexsym}
\usepackage{graphicx,color}
\usepackage{caption}
\usepackage{float}
\usepackage{slashed}
\usepackage{graphicx}
\graphicspath{ {images/} }
\usepackage{ytableau}

\usepackage{multirow}

\usepackage{mathtools}

\usepackage[colorinlistoftodos]{todonotes}
\usepackage[colorlinks=true, allcolors=blue]{hyperref}

\usepackage{wrapfig}
	\usepackage[T1]{fontenc}
	\usepackage{tikz}
	\usetikzlibrary{decorations.pathmorphing}
\usepackage{tikz}
\usetikzlibrary{shapes.geometric, arrows,patterns,snakes}
\tikzstyle{ellip} = [ellipse, minimum width=3cm, minimum height=1cm,text centered, draw=black]

\newskip\humongous \humongous=0pt plus 1000pt minus 1000pt

\newif\ifdtup

\allowdisplaybreaks[1]

\catcode`\@=11

\@addtoreset{equation}{section}

\def\@normalsize{\@setsize\normalsize{15pt}\xiipt\@xiipt
\abovedisplayskip 14pt plus3pt minus3pt%
\belowdisplayskip \abovedisplayskip
\abovedisplayshortskip \z@ plus3pt%
\belowdisplayshortskip 7pt plus3.5pt minus0pt}

\def\small{\@setsize\small{13.6pt}\xipt\@xipt
\abovedisplayskip 13pt plus3pt minus3pt%
\belowdisplayskip \abovedisplayskip
\abovedisplayshortskip \z@ plus3pt%
\belowdisplayshortskip 7pt plus3.5pt minus0pt
\def\@listi{\parsep 4.5pt plus 2pt minus 1pt
     \itemsep \parsep
     \topsep 9pt plus 3pt minus 3pt}}

\relax

\catcode`@=12

\catcode`\@=11

\def\section{\@startsection{section}{1}{\z@}{3.5ex plus 1ex minus
   .2ex}{2.3ex plus .2ex}{\large\bf}}

\def\SymBoxes#1#2#3#4{\newdimen\un@t \un@t#3%
\raisebox{#1}{\rule{#2\un@t}{#4}\hskip-#2\un@t
\@tempdimb\un@t \advance\@tempdimb by-#4\@tempcntb#2\relax%
\@whilenum{\@tempcntb>0}\do{
\rule{#4}{\un@t}\hskip\@tempdimb \advance\@tempcntb by\m@ne}%
\hskip-#2\un@t \rule[\un@t]{#2\un@t}{#4}%
\rule[\un@t]{#4}{#4}\hskip-#4
\rule{#4}{\un@t}}\hskip-#4}                

\begin{document}

\newcommand{\beq}{\begin{equation}}
\newcommand{\eeq}{\end{equation}}
\newcommand{\bea}{\begin{eqnarray}}
\newcommand{\eea}{\end{eqnarray}}
\newcommand{\beas}{\begin{eqnarray*}}
\newcommand{\eeas}{\end{eqnarray*}}
\newcommand{\defi}{\stackrel{\rm def}{=}}
\newcommand{\non}{\nonumber}
\newcommand{\bquo}{\begin{quote}}
\newcommand{\enqu}{\end{quote}}
\renewcommand{\(}{\begin{equation}}
\renewcommand{\)}{\end{equation}}
\def \eqn#1#2{\begin{equation}#2\label{#1}\end{equation}}
\def\IZ{{\mathbb Z}}
\def\IR{{\mathbb R}}
\def\IC{{\mathbb C}}
\def\IQ{{\mathbb Q}}
\def\de{\partial}
\def\Tr{ \hbox{\rm Tr}}
\def\H{ \hbox{\rm H}}
\def\HE{ \hbox{$\rm H^{even}$}}
\def\HO{ \hbox{$\rm H^{odd}$}}
\def\K{ \hbox{\rm K}}
\def\Im{ \hbox{\rm Im}}
\def\Ker{ \hbox{\rm Ker}}
\def\const{\hbox {\rm const.}}
\def\o{\over}
\def\im{\hbox{\rm Im}}
\def\re{\hbox{\rm Re}}
\def\bra{\langle}\def\ket{\rangle}
\def\Arg{\hbox {\rm Arg}}
\def\Re{\hbox {\rm Re}}
\def\Im{\hbox {\rm Im}}
\def\exo{\hbox {\rm exp}}
\def\diag{\hbox{\rm diag}}
\def\longvert{{\rule[-2mm]{0.1mm}{7mm}}\,}
\def\a{\alpha}
\def\dag{{}^{\dagger}}
\def\tq{{\widetilde q}}
\def\p{{}^{\prime}}
\def\W{W}
\def\N{{\cal N}}
\def\hsp{,\hspace{.7cm}}

\def\br{\nonumber\\}
\def\IZ{{\mathbb Z}}
\def\IR{{\mathbb R}}
\def\IC{{\mathbb C}}
\def\IQ{{\mathbb Q}}
\def\IP{{\mathbb P}}
\def \eqn#1#2{\begin{equation}#2\label{#1}\end{equation}}

\newcommand{\sgm}[1]{\sigma_{#1}}
\newcommand{\idd}{\mathbf{1}}

\newcommand{\C}{\ensuremath{\mathbb C}}
\newcommand{\Z}{\ensuremath{\mathbb Z}}
\newcommand{\R}{\ensuremath{\mathbb R}}
\newcommand{\rp}{\ensuremath{\mathbb {RP}}}
\newcommand{\cp}{\ensuremath{\mathbb {CP}}}
\newcommand{\vac}{\ensuremath{|0\rangle}}
\newcommand{\vact}{\ensuremath{|00\rangle}                    }
\newcommand{\oc}{\ensuremath{\overline{c}}}
\begin{titlepage}
\begin{flushright}
CHEP XXXXX
\end{flushright}
\bigskip
\def\thefootnote{\fnsymbol{footnote}}

\begin{center}
{\Large
{\bf Bounds on Slow Roll and the de Sitter Swampland
\vspace{0.2in}
}
}
\end{center}

\bigskip
\begin{center}
{Sumit K. GARG$^a$,\footnote{\texttt{sumit.k@cmr.edu.in}} \ \ Chethan KRISHNAN$^b$\footnote{\texttt{chethan.krishnan@gmail.com}} \vspace{0.15in} \\ }
\vspace{0.1in}

\end{center}

\renewcommand{\thefootnote}{\arabic{footnote}}

\begin{center}
$^a$ {Department of Physics, \\
CMR University, Bengaluru 562149, India }

$^b$ {Center for High Energy Physics,\\
Indian Institute of Science, Bangalore 560012, India}

\end{center}

\noindent
\begin{center} {\bf Abstract} \end{center}
The recently introduced swampland criterion for de Sitter (arXiv:1806.08362) can be viewed as a (hierarchically large) bound on the smallness  of the slow roll parameter $\epsilon_V$. This leads us to consider the other slow roll parameter $\eta_V$ more closely, and we are  lead to conjecture that the bound is not necessarily on $\epsilon_V$, but on slow roll itself. A natural refinement of the de Sitter swampland conjecture is therefore that slow roll is violated at ${\cal O}(1)$ in Planck units in any UV complete theory. A corollary is  that $\epsilon_V$ need not necesarily be ${\cal O}(1)$, if $\eta_V \lesssim -{\cal O}(1)$ holds. We consider various tachyonic tree level constructions of de Sitter in IIA/IIB string theory (as well as closely related models of inflation), which superficially violate arXiv:1806.08362, and show that they are consistent with this refined version of the bound. The phrasing in terms of slow roll makes it plausible why both versions of the conjecture run into trouble when the number of e-folds during inflation is high. We speculate that one way to evade the bound could be to have a large number of fields, like in $N$-flation. 




\vspace{1.6 cm}
\vfill

\end{titlepage}

\setcounter{page}{2}

\setcounter{footnote}{0}



\section{De Sitter Denialism}

Constructing fully trustable de Sitter vacua in string theory has turned out to be a hard problem. There are various no-go theorems in effect which make de Sitter impossible at the level of classical 10 and 11 dimensional supergravity \cite{Carlos}, and there have been multiple claims in the recent literature that there is no fully reliable construction even at the level of the full string theory \cite{Ulf, VafaReview}. One particularly damning criticism (if true) is perhaps the claim by Sethi \cite{Sethi} that IIB non-supersymmetric AdS flux vacua of \cite{GKP} are not reliable starting points for uplift to de Sitter\footnote{Let us give a mini-summary of the situation here for the general population. There are many known ways to construct de Sitter ``string" vacua if one's starting point is a four dimensional effective supergravity, with potentials that are well-motivated (but not fully explicitly derivable) from ten dimensions \cite{many}. This is where the string landscape has its origins. The trouble is in deriving these models from the point of view of the 10/11 dimensional string/M theory, with calculations that are fully under control. In the 10 dimensional setting, all known constructions of de Sitter rely on indirect arguments (like existance of non-perturbative corrections, non-geometric fluxes, etc) where not every piece in the construction is {\em explicitly and reliably} calculable. In set ups where things are reliably calculable (these are sometimes called classical or tree level constructions, even though they contain D-branes and O-planes), one either does not find de Sitter, or when one finds it, it is  tachyonic. One of the claims in this paper is that these tachyonic directions are necessarily steep: the tachyons are Planck scale.}. 

We will take the stand in this paper that string theory does not admit de Sitter vacua\footnote{By the end of this paper, we will see that the arguments presented in \cite{Vafa1} and in this paper are really evidence against tree level constructions with sources. But the de Sitter swampland conjecture is  made more generally.}, not because we believe this is an inevitable conclusion yet, but because we do believe it could be a useful exploratory schtick. We will focus on a set of ideas that has emerged recently as a result of the concern about de Sitter in string theory. The basic suggestion \cite{Vafa1} is to elevate the inability to find de Sitter into a feature rather than a problem, and to declare (in a suitable way that we will make precise) that de Sitter must be absent in any consistent theory of quantum gravity. The way this is done is by first noting that in string theory compatifications, the cosmological constant is realized as the value of a minimum of a scalar potential. The idea that there are no de Sitter solutions in a consistent quantum gravity is then translated into a statement about that scalar potential. In particular, the bound that was proposed in \cite{Vafa1} takes the form
\bea
|\nabla V| \ge c \ V \label{1stbound}
\eea
The (positive) number $c$ was not given a universal value, but it is expected to be an ${\cal O}(1)$ number in Planck units. In other words, when the Planck mass goes to infinity, and we are working with effective effective theory without worrying about the UV completion, the bound becomes a tautology. This means, at its strongest, that inflation and de Sitter can exist only in effective field theory, but not in a consistent quantum theory of gravity\footnote{Indeed, meta-stable supersymmetry breaking and positive vacuum energy in field theory without gravity is likely generic \cite{ISS}. It is also a cosmologically viable mechanism for supersymmetry breaking \cite{CK}.}. Note also that in supersymmetric vacua, the right hand side is zero or negative, and the bound is automatic.

The form of the inequality above is one that has come up previously in discussions on the existence of de Sitter/inflating vacua in string theory, eg., \cite{HKTT, Timm}, in attempting to constrain de Sitter/inflating models, eg., in Type IIA string theory. It was shown in these papers by demonstrating an inequality of the above form, that various stringy constructions could not give rise to inflation/de Sitter. What was done in \cite{Vafa1}, was that these observations were elevated into a general principle, while giving a scan of various models. 

Even though it was not particularly emphasized in \cite{Vafa1}, it is clear that this condition can be viewed as a statement about the ``slow roll'' parameter $\epsilon_V$ in quantum gravity\footnote{This refers to what is called potential slow roll, to be distinguised from Hubble slow roll.}. The bound above is the statement that $\epsilon_V$ must satisfy the inequality
\bea
\epsilon_V \equiv \frac{1}{2}\frac{(\nabla V)^2}{V^2} \ge \frac{c^2}{2}. \label{secondbound}
\eea
Viewed this way, we are naturally lead to conjecture that it is not merely the parameter $\epsilon_V$ that is violated at ${\cal O}(1)$ in quantum gravity, but the idea of slow roll itself. This is the idea we will investigate in this paper. A necessary condition for slow roll\footnote{This is strictly true only for single field slow roll. See \cite{future} for further discussions on some related questions.} is that one must have not just $\epsilon_V \ll 1$, but also that $|\eta_V | \ll 1$, where
\bea
\eta_V \sim \frac{\nabla^2 V}{V}.
\eea
The two inequalities 
\bea
\epsilon_V \ll 1, \ \ {\rm and} \ \  |\eta_V | \ll 1
\eea
are together called the Slow Roll Conditions. The second condition is necessary in conventional inflationary models to make sure that $\epsilon_V$ remains small for enough number e-folds. In other words, loosely the demand of inflation is not just that $\epsilon_V$ is small, but that it is stably small. 

The slow roll conditions {\em must} hold for inflation to happen, but they might not be sufficient. In this paper however, we will treat slow roll as equivalent to inflation\footnote{More precisely, our statements will be about slow roll and not about inflation directly, if the reader wishes to make a distinction between the two.}. This is partly justified by the fact that inflation is of interest typically around attractor solutions, where the two conditions are indeed {\em also} sufficient \cite{LyddleLyth}. In any event, once we make this equivalence, it follows that to rule out inflation we do not necessarily need to rule out $\epsilon_V \ll 1$: we just need to break one of the two slow roll conditions, and have either $\epsilon_V \ll 1$ or $|\eta_V | \ll 1$ violated at ${\cal O}(1)$. This is the main observation we make in this note, we will present various pieces of evidence in string theory, for this refinement. The refined de Sitter swampland criterion then is that {\it at least one of $\epsilon_V$ or $|\eta_V|$ must be bigger than ${\cal O}(1)$ in string theory}\footnote{We will discuss only those cases in this paper, where we are near a critical point, so that we can get a clean bound on $\eta_V$. It will also be interesting to have a statement about the situation where $\epsilon_V$ and $\eta_V$ are intermediate, we will present some results on this in \cite{future}.}. Our work can be viewed as an investigation of the case one can make for de Sitter denialism in string theory\footnote{Our technical discussion, like that in \cite{Vafa1} will be in the context of tree level constructions however.}. We will demand these conjectures here only for states with positive potential energy, to avoid dealing with (possibly supersymmetric) Minkowski and AdS vacua. In \cite{Vafa1} this same goal was accomplished by working with \eqref{1stbound} rather than \eqref{secondbound}.


Obviously, all the cases that were investigated in \cite{Vafa1} and were found to be consistent with their conjecture, will go through in our case as well. Interestingly, even though it was not discussed in \cite{Vafa1}, there exist results in the literature which are either in tension or superficially in violation of the conjecture in \cite{Vafa1}. 
These are the cases corresponding to ``tree level'' de Sitter constructions in type II string theory, which have been investigated quite a bit. These solutions are known to be tachyonic \cite{Ulf}. Some of these solutions in fact do contain solutions that naively violate the bound of \cite{Vafa1}. We will argue that when the conjecture of \cite{Vafa1} is refined as we have suggested above, those results also become consistent with the general claim that slow roll should be vioated at ${\cal O}(1)$ in Planck  units. 

To do this, we take advantage of the fact that the $\eta_V$-parameter is related to the eigenvalues of the second derivative matrix of the scalar potential. Let us also note that the approach we use to put bounds on (tachyonic) eigenvalues  might be of some interest in itself. We suspect that it could direct our intuition about what is the best approach (eg., the choice of sources and cycles they wrap, internal geometry, etc) if one hopes to beat these bounds in tree level string theory. 

{\it Added note:} A few papers which are relevant to this topic have appeared since \cite{Vafa1}, including \cite{Vafa2, Andriot, Ulfnew, Bert, Achucarro}. The paper \cite{Andriot} was the first to note the tension of the \cite{Vafa1} conjecture with the existence of tachyonic de Sitter constructions, it also suggested a few different ways to evade the tension. Note that our aim here is different: we wish to give a {\em bound} for the tachyons in the spirit of \cite{Vafa1}. What is surprising is that the bound holds in all cases we have checked.


\section{Slow Roll in Type II String Theory}

We will work with type II compactifications \cite{HKTT, Shiu} to 4 dimensions with the following reduction ansatz
\bea
ds_{10}^2=\tau^{-2}ds_4^2+\rho ds_6^2
\eea
where $\rho$ captures the volume modulus of the compact space and $\tau = e^{-\phi} \rho^{3/2}$ with $\phi$ the dialton. The choice is dictated by the demand that these universal moduli do not mix in the 4D Einstein frame. It turns out that if one defines (we set the reduced Planck mass to one, throughout this paper)
\bea
\hat \rho \equiv \sqrt{3/2} \ \ln \rho,  \ \ \hat \tau \equiv \sqrt{2} \ln \tau
\eea
the kinetic terms of these fields become canonically normalized \cite{HKTT}. The remarkable fact is that we can make fairly general statements about the nature of de Sitter vacua in string theory even while we restrict our attention to the $\rho$-$\tau$ subspace of the moduli space. We will consider a fairly generic scenario where potentials in 4 dimensions arise from curvature in the the compact space, NSNS fluxes $H_3$, RR fluxes $F_p$ and localized brane and orientifold ($D_q/O_q$) sources.
\bea
V_{R_6}=\frac{A_{R_6}}{\tau^2\rho}, \ \ V_{H_3}=\frac{A_{H_3}}{\tau^2 \rho^3}, \ \ V_{F_p}=\frac{A_{F_p}}{\tau^4\rho^{(p-3)}}, \ \ V_q=\frac{A_{q}}{\tau^3\rho^{(6-q)/2}}
\eea
We will follow the discussion in \cite{Shiu} for concreteness, and because the set of cases there is sufficiently rich to illustrate the points we wish to make. In particular, they are general enough to give rise to the types of situations that are covered in \cite{Vafa1}. 

\subsection{Shiu-Sumitomo Tachyons}

The key observation made in \cite{Vafa1} is that this generic set up is sufficient to put bounds on the slow roll parameter $\epsilon_V$ in many (but not all) situations. Nonetheless, it is worth noting that there exist resuts (eg., \cite{Sonia, Koerber}) in the literature that construct de Sitter solutions in explicit tree level constructions, that do seem to superficially violate the bound suggested in \cite{Vafa1}. Our de Sitter denialism based conjecture is that such cases will always be tachyonic, and that the corresponding $\eta_V$-parameter will be ${\cal O}(1)$ in Planck units. When the tachyons are not in the $\rho$-$\tau$ plane, we cannot say much about them from the above general set up. But when the tachyons do lie in that plane, we can check if they satisfy our bound\footnote{Note that it is not {\em necessary} that they satisfy the bound: the slow roll parameter $\eta_V$ is defined via the smallest eigenvalue of the second derivative matrix of potentials, and the smallest eigenvalue could be in some other direction in field space. But it is a {\em sufficient} condition, so if the smallest eigenvalue in this subspace already satisfies the bound, the full set of moduli also necessarily satisfies the bound. Remarkably, in {\em every} case that we have investigated, whenever there is a tachyon in the $\rho$-$\tau$ plane, it already satisfies the bound.}. 

The straightforward way to do this is to compute the eigenvalues of the second derivatives at the critical points. But calculating the critical values and the values of the moduli at them explicitly is often forbiddingly complicated even for simple combinations of potentials because of the various combinations of powers involved. In \cite{Shiu} the goal was to merely show that the tachyons {\em existed}, so they could get some mileage via a frontal attack. But what we wish to show is that the tachnyonic eigenvalues are bounded, so this will not work. Instead, we will use the fact that there are various positivity inequalities that the sources have to satisfy, together with the fact that the critical point can be defined via 
\bea
\rho \frac{\partial V_{eff}}{\partial \rho}=0, \ \ \tau \frac{\partial V_{eff}}{\partial \tau}=0.
\eea
The latter expressions are written in a form that enables us to take advantage of the homogeneity properties, and therefore these relations can be re-expressed in terms of the various potentials. This allows us to bypass dealing with the complications of the moduli directly. 

Let us present one case in detail to show the mechanics of the approach. We have checked our method for all the cases presented in section 3 of \cite{Shiu}.

\subsection{IIA: ($R_6, H_3, F_0, F_2, O_4$)}

Together with the NS-Ns flux and the two RR-fluxes, we also have an $O4$-plane source.
\begin{eqnarray}
V_{eff} &=& V_{R_{6}} + V_{H_{3}} + V_{F_{0}}+ V_{F_{2}}-V_{O4}\nonumber\\
&=& \frac{A_{R_6}}{\tau^2\rho} + \frac{A_{H_{3}}}{\tau^2 \rho^3} + \frac{A_{F_{0}}}{\tau^4 \rho^{-3}}+ \frac{A_{F_{2}}}{\tau^4 \rho^{-1}}
-\frac{A_{O4}}{\rho\tau^3 }
\end{eqnarray}
Minimization Conditions $\rho \frac{\partial V_{eff}}{\partial \rho}=0$ and $\tau \frac{\partial V_{eff}}{\partial \tau}=0$  give
\begin{itemize}
 \item $V_{H_{3}}=2 V_{eff}|_{extr}-V_{R_6}+ \frac{1}{2}V_{O4}$
 \item $V_{F_{0}}=\frac{7}{2}V_{eff}|_{extr}-V_{R_6}$
 \item $V_{F_{2}}=-\frac{9}{2}V_{eff}|_{extr}+ V_{R_6}+ \frac{1}{2}V_{O4}$
\end{itemize}
The canonically normalized fields are \[ \hat{\rho} = \sqrt{\frac{3}{2}}~\ln\rho, ~ \hat{\tau}=\sqrt{2}~\ln\tau\]
They yield the matrix elements of the (symmetric) second derivative matrix $\partial_i \partial_j V_{eff}$  of the potential to be
\begin{eqnarray}
  M_{11} &=& \frac{\partial^2 V_{eff}}{\partial \hat{\rho}^2} = \frac{2}{3}(9 V_{H_{3}} + V_{R_{6}} + 9 V_{F_{0}}+ V_{F_{2}}-V_{O_{4}}) \nonumber\\
 M_{22} &=& \frac{\partial^2 V_{eff}}{\partial \hat{\tau}^2}= \frac{1}{2}(4 V_{H_{3}} +4 V_{R_{6}} +16 V_{F_{0}}+ 16 V_{F_{2}}-9 V_{O_{4}}) \nonumber\\
 M_{12} &=& \frac{\partial}{\partial \hat{\tau}}(\frac{\partial V_{eff}}{\partial \hat{\rho}})=\frac{1}{\sqrt{3}}(6 V_{H_{3}}+2 V_{R_{6}} -12 V_{F_{0}}-4 V_{F_{2}}
  -3 V_{O_{4}}) 
\end{eqnarray}
Now the smaller of the eigenvalues can be obtained as
\bea
m^2_T=\frac{1}{2} \left( {\rm Tr}[M]_{extr} - ({\rm Tr}[M]^2-4 {\rm Det}[M])^{1/2}_{extr}\right) \label{tachmass}
\eea
where
\begin{eqnarray}
  {\rm Tr}[M]_{extr} &=& 26 V_{eff} +\frac{19}{6} V_{O_{4}}-\frac{32}{3} V_{R_{6}}\nonumber\\
  ({\rm Tr}[M]^2-4 {\rm Det}[M])_{extr}&=& 1348 V_{eff}^2 +\frac{361}{36} V_{O_{4}}^2 + \frac{1216}{9} V_{R_{6}}^2 +\frac{634}{3} V_{eff} V_{O_{4}}\nonumber\\
 && -\frac{2560}{3} V_{eff} V_{R_{6}}-\frac{608}{9} V_{O_{4}} V_{R_{6}}
\end{eqnarray}
Note that throughout this procedure our key strategy is to express the extremum condition and the second derivatives at the extremum via the values of the potentials at the extremum. This is possible because we can take advantage of their scaling properties. 

Finally, to obtain the bound, we note that $V_{H_3},  V_{F_{0}}, V_{F_{2}}$ have to be non-negative, and this puts positivity conditions on the right hand sides of the bulleted equations above. The range of values allowed by these inequalities for $x\equiv V_{O_4}/V_{eff}$ and $y \equiv V_{R_6}/V_{eff}$ are plotted in  figure 1.
 
\begin{figure}[h]\centering
\hspace{-5mm}
\includegraphics[angle=0,width=80mm]{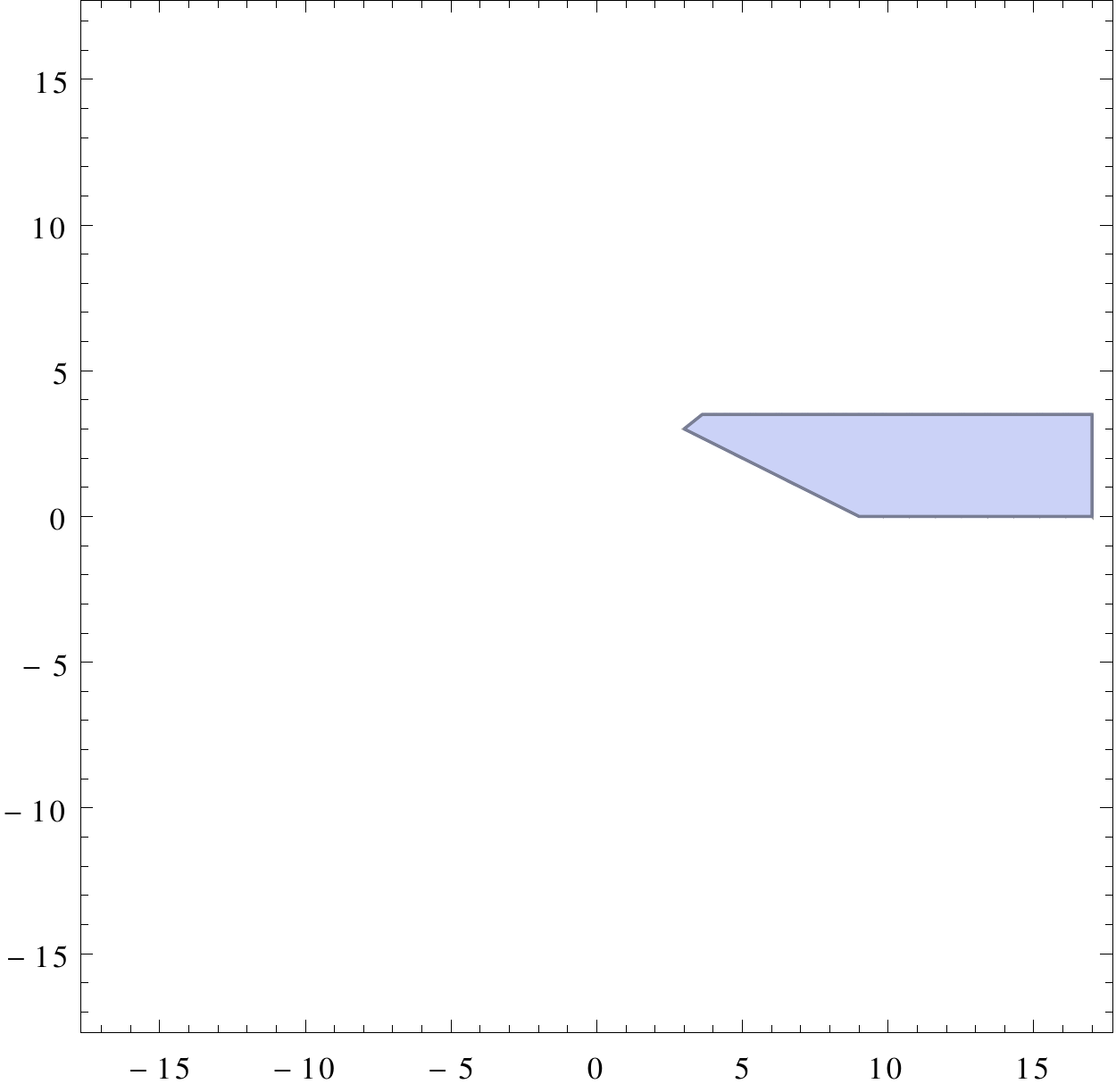} \\
\caption{Plot of allowed values of $x\equiv V_{O_4}/V_{eff}$ and $y \equiv V_{R_6}/V_{eff}$, the region extends indefinitely to the right.}
\label{fig13R}
\end{figure}

Remarkably, it turns out that if one scans \eqref{tachmass} in this region, we find an upper bound on 
\bea
\eta_V \equiv \frac{m_T^2}{V_{eff}} \approx -3.68
\eea
It should be possible to find analytic expressions for these quantities since we know the domain where the function needs to be evaluated. But since these bounds are very crude to begin with, we will not attempt to do so. A full list of all the bounds is provided in the table in the next subsection, and some of the details of the various cases is provided in an Appendix.

\subsection{A Higher Dimensional Tachyon of Van Riet}

Together with the various 4 dimensional cases presented in \cite{Shiu}, let us also look at one higher dimensional case we are aware of, due to Van Riet \cite{VanRiet}. This is a IIA set up in 7D\footnote{See \cite{VanRiet2} for a related T-dual version in IIB.}, with $O6$ sources, and it reveals one interesting point. In the construction in \cite{VanRiet} the flux $F_2$ is assumed to be present, but we do {\em not} find a bound when it is present. We believe this is because in a genuine string theory construction in seven dimensions, the $F_2$ will not survive the orientifold projection (see the discussion in section (3.2) in \cite{VanRiet}). Indeed, when we drop the $F_2$ source, we do find a bound, as we show below. The approach is parallel to the previous subsection, so our presentation is somewhat telegraphic. 
\begin{eqnarray}
V_{eff} &=&  V_{F_{0}}- V_{O} +V_{H_{3}}+ V_R\nonumber\\
&=& \frac{A_{F_{0}}}{\tau^7 \rho^{-3/2}} - \frac{A_{O}}{\tau^{9/2} \rho^{3/4}}+ \frac{A_{H_{3}}}{\tau^2\rho^3 }
+\frac{A_{R}}{\tau^2 \rho }
\end{eqnarray}
Minimization Conditions yield
\begin{itemize}
 \item $V_{R}=\frac{12}{5} V_{eff}|_{extr}$
 \item $V_{F_{0}}= \frac{3}{5} V_{eff}|_{extr}+ V_{H_{3}}$
 \item $V_{O}= 2 V_{eff}|_{extr}+ 2 V_{H_{3}}$
\end{itemize}
We define quantities\footnote{We are not being very careful with the precise proportionality coefficients relating us to the canonically normalized fields. In 4D this only affects the value of the bound by an ${\cal O}(1)$ number, not the existence of the bound itself.}  
\[ \hat{\rho} \propto  \ln\rho, ~ \hat{\tau}\propto \ln\tau\]
The matrix elements are given by 
\begin{eqnarray}
  M_{11} &=& \frac{\partial^2 V_{eff}}{\partial \hat{\rho}^2} = \frac{9}{4} V_{F_{0}}- \frac{9}{16}V_{O} +9 V_{H_{3}}+V_R\nonumber\\
 M_{22} &=& \frac{\partial^2 V_{eff}}{\partial \hat{\tau}^2}= 49 V_{F_{0}}- \frac{81}{4}V_{O} +4 V_{H_{3}}+4 V_R \nonumber\\
 M_{12} &=& \frac{\partial}{\partial \hat{\tau}}(\frac{\partial V_{eff}}{\partial \hat{\rho}})=-\frac{21}{2} V_{F_{0}}
 - \frac{27}{8}V_{O} +6 V_{H_{3}}+ 2 V_R
\end{eqnarray}
yielding  
\begin{eqnarray}
  {\rm Tr}[M]_{extr} &=& \frac{9}{8} V_{eff}+\frac{181}{9} V_{H_{3}}\nonumber\\
  ({\rm Tr}[M]^2-4 {\rm Det}[M])_{extr}&=& \frac{18513}{64} V_{eff}^2 + \frac{32761}{64} V_{H_{3}}^2 + \frac{23133}{32} V_{eff} V_{H_{3}}\nonumber\\
 \end{eqnarray}
Following the approach as before, we find that there is a maximum value
\bea
\eta_V \le -7.94
\eea
Some details of the various remaining cases are presented in the Appendix, here we summarize the bounds of all the cases we have looked at.\\

{\renewcommand{\arraystretch}{1.9}
 \begin{tabular}{|p{4.2cm}|p{4.4cm}|p{2.2cm}|}
\hline
\multicolumn{3}{|c|}{Bound Summary } \\
\hline
Different Scenarios & Ingredients & Bound \\ \hline
\multirow{2}{*}{IIA in 4D} & $\{ R_{6},~H_{3},~F_{0},~F_{2},~O_{4}\}$ & $\eta_V\leq -3.68$ \\\cline{2-3}
 & $\{ R_{6},~H_{3},~F_{0},~O_{4},~O_{6}\}$ & $\eta_V\leq -3.23$ \\ \hline\hline
\multirow{4}{*}{IIB in 4D}
& $\{ R_{6},~H_{3},~F_{1},~F_{3},~O_{3}\}$  & $\eta_{V}\leq -4.0$  \\\cline{2-3}
 &$\{ R_{6},~H_{3},~F_{1},~O_{3},~O_{5}\}$  & $\eta_{V}\leq -3.42$ \\\cline{2-3}
 & $\{ R_{6},~H_{3},~F_{1},~O_{3},~O_{7}\}$  & $\eta_{V}\leq -4.0$  \\\cline{2-3}
 & $\{ R_{6},~H_{3},~F_{3},~O_{3},~O_{7}\}$  & $\eta_{V}\leq -4.0$  \\\cline{2-3}
 \hline\hline
IIA in 7D & $\{ R_{6},~H_{3},~F_{0},~O_{6}\}$ & $\eta_V\leq -7.94$  \\ \hline
\end{tabular}

\section{Tachyons in Explicit Type II Compactifications}

The cases that we have discussed so far are based on powerful general arguments on the volume-dilaton subspace of the moduli space. The cases where tachyons exist there, we have shown that those tachyons satisfy our bound. But while these arguments are very powerful where they work, they suffer from a drawback: tachyons need not live in the volume-dilaton subspace. Nonetheless, it is known that all the tree level constructions of de Sitter in Type II string theory to date contain tachyons, so it behooves us to check that they are consistent with our bounds. To the extent that we have been able to check the literature, this indeed seems to be the case. We will present some of the details in this section.

In particular, let us first consider the IIA models with metric fluxes on toroidal orientifolds which were discussed in \cite{Sonia}. In most such cases, a small $\epsilon_V$ is ruled out by no-go theorems, but they manage to find some classes of solutions that allow a slow roll paramter $\epsilon_V \approx 0$\footnote{Note that our discussions in this paper are always either at the critical point or very close to it. This is because then we can aim to get an unambiguous bound on  $\eta_V$. It will be interesting to generalize our discussion to include the constructions in \cite{Dibitetto} which contain solutions with $\epsilon_V \sim 0.4$ and $\eta_V \sim -0.1$, which undergo a few e-foldings. Some work in this direction will be reported elsewhere.}  for some classes of orientifolds of $\IZ_2 \times \IZ_2$. But in {\em all} such cases that they discuss and numerically construct, the numerical values of the tachyons are quoted as $\eta_V \lesssim -2.4$, immediately consistent with our bound. In many other cases that have been discussed in the literature, we have not been able to find explicit quotes of the tachyon mass, except for the general statement that they do exist. Because the explicit constructions often involve numerical solutions of complicated polynomial type equations, not much more can be said by a direct scan of the literature. It will be interesting to do a scan of the existant solutions and their tachyons to see if they all have to satisfy an ${\cal O}(1)$ bound\footnote{In particular, an especially interesting possibility is to know whether the bound can be relaxed to ${\cal O}(1)/N$ if there are $N$ tachyons. But from the discussions in the literature that we could find, whether this possibility is allowed is not clear to us. The number of tachyons in the literature is at most $\lesssim {\cal O}(10)$.}. In \cite{Koerber}, a consolidation of the tachyon situation in the literature is presented, and the tachyons they found all have  $\eta_V \lesssim -2 $, again in agreement with our bound. The orbifold groups in these cases always contain $\IZ_2 \times \IZ_2$, so it is not clear (to us) that the results in this regard in \cite{Koerber} are truly distinct from those in \cite{Sonia}.

Finally, a very suggestive calculation has been done in a top-down motivated 4D set up in \cite{Zagermann}. They identified a universal class of tachyons for de Sitter solutions of ${\mathcal N}=1$ supergravity with an F-term potential near a no-scale Minkowski vacuum. They find that this entire class of tachyons satisfy the bound $\eta_V \le -4/3$, which again is entirely consistent with our bound.

\section{De Sitter Denialism, Again}

We will discuss some pros and cons of de Sitter denialism in this section in lieu of a conclusion. 

There are two immediate challenges to the conjecture made in \cite{Vafa1} and its refinement that we have presented here. Firstly, there is the observational evidence that dark energy is a fluid with equation of state $w \approx -1$. This comes not just from WMAP, but also from baryon acoustic oscillations and the two are strikingly consistent. This does not mean that dark energy is necessarily a cosmological constant, but it does raise a question for quintessence afficianados why the equation of state seems close to $w=-1$, when it did not have to be. The second issue has to do with inflation in the early universe. It was pointed out in \cite{Vafa2} that the many e-foldings required by the inflationary scenario to solve the horizon, flatness and related problems is in tension with the bound presented in \cite{Vafa1}. Our discussion in this paper clarifies this point by noting that the bound should be understood as a bound on single field slow roll itself, and therefore a high number of e-folds is pretty much in direct contradiction with it\footnote{This is true in single field inflation, but requires further comments in multi-field slow roll \cite{Achucarro, future}.}.

Inflation in many situations is known to have problems due to trans-Planckian field ranges during slow roll, and it has often been suggested (eg.,\cite{Nflation, Achucarro}) that one way to avoid this is to have a large number of (axionic?) fields like in N-flation. It is conceivable that the field range bound \cite{Vafa2}, which says $\Delta \phi \lesssim {\cal O}(1)$ together with the slow roll bounds might be evaded if we have a large number of fields. It will be interesting to see if one can generate heirarchically small slow roll parameters in this approach, especially since string theory has many light scalar fields.

Our claims in this paper put the old observations about challenges to constructing de Sitter in string theory \cite{Willy, Susskind}, like the $\eta$-problem, together with the more recent observations about the inability to construct non-tachyonic de Sitter in tree level type II constructions \cite{Ulf}, as well as the conjectures in \cite{Vafa1, Vafa2} into a somewhat unified setting. The extra ingredient is that we claim that the tachyonic instability should be Planck scale, ie., that the violation of $\eta_V$-slow roll is $ {\cal O}(1)$ in Planck mass. A bit surprisingly, we were able to show using general scaling-type arguments that the known tachyons in type II tree level constructions fit this conjecture. The previous work which tried to study the nature of these tachyons \cite{Shiu} tried to get a handle on them by working directly with the moduli at the de Sitter critical point. Our strategy has been different and was more in the spirit of \cite{HKTT} and \cite{Vafa1}. We found evidence for our claims by expressing the relevant quantities in terms of the potential contributions themselves and not directly in terms of the moduli, because the latter get complicated very quickly. This enabled us to not just note that there are tachyons, but also to put bounds on their (tachyonic) masses.

One way to view the conjecture in \cite{Vafa1} is as an indication that one has to come up with a new paradigm for dark energy and inflation in a string theoretic context. See \cite{Ulfnew} for a very recent attempt in this direction. But before we conclude, we will make some subversive comments that could be taken as evidence that the conjecture of \cite{Vafa1} (and this paper) should be viewed only as a statement constraining tree level string constructions of de Sitter, and not the full string theory.

There are two main motiations we can see for the suggestion in \cite{Vafa1}. One is that there are no fully explicit   constructions of de Sitter in string theory, even with quantum effects included. Second is that (as we have emphasized many times) all the explicit, but classical, constructions of de Sitter in string theory are tachyonic. Note that these points are perhaps unsurprising, and might be a feature of de Sitter than a flaw of string theory. This is because unlike AdS, which can be constructed in flat space by putting branes, a de Sitter space can only be constructed in flat space via something akin to a Big Bang singularity -- this is a result of Penrose's singularity theorem forbidding de Sitter in asymptotically flat classical general relativity, and was noted in the classic paper of Farhi and Guth \cite{Farhi-Guth}. Given this fact, we do not find it particularly surprising that de Sitter constructions in string theory require quantum corrections: indeed the whole point of the Farhi-Guth like argumentation is in effect, that one cannot build de Sitter classically. The real trouble in our view is that no fully explicit (and unambiguosly controlled) constructions exist, {\em not} that the constructions are not classical\footnote{It is also stated sometimes as a cause of worry, that {\em none} of the known constructions, in any duality frame, are fully classical. This again, we find to be a consequence of the expectation that classical effects alone cannot generate de Sitter. Note that schematically, a duality rotation on ``classical+quantum'' must give us ``quantum+classical'', and not {\em just} "classical".}. 

One counter-argument to our counter-argument in the last paragraph is that even though the tree level type II constructions are often called classical, they do contain localized brane sources etc., which capture many (though not all) non-perturbative effects. So perhaps the fact that we have not been able to construct a de Sitter vacuum without tachyons in this context, should indeed cause us worry.  


\section*{Acknowledgments}

SKG thanks the Center for High Energy Physics (IISc) and B. Ananthanarayan, for hospitality. CK thanks the organizers of Strings 2018, Okinawa, for a stimulating conference: this paper is a direct result of C. Vafa's talk. We thank David Andriot, Ulf Danielsson, Giuseppe Dibitetto, and especially Thomas Van Riet for comments on a previous version of the paper. 


\appendix

\section{Bounding via Scaling}

Here we present the details of all the generic tachyons in the volume-dilaton plane discussed in \cite{Shiu} and how we can bound their (squared) masses\footnote{Our aim here is to demonstrate the power of our strategy and that it yields the ${\cal O}(1)$ bounds that we seek. Some of these cases can be further constrained using other No-Go theorems, see eg. \cite{Andriot2} and references therein.}. We do not consider the cases they consider that do not contain tachyons (or contain unfixed moduli). One case was already presented in detail in the main body of the text, so here we will just present the key equations of the remaining cases. The bounds are collected in Table 1.

\subsection{IIA: $R_6, H_3, F_0, O_4, O_6$}
\begin{eqnarray}
V_{eff} &=& V_{R_{6}} + V_{H_{3}} + V_{F_{0}}- V_{O_{4}}-V_{O6}\nonumber\\
&=& \frac{A_{R_6}}{\tau^2\rho} + \frac{A_{H_{3}}}{\tau^2 \rho^3} + \frac{A_{F_{0}}}{\tau^4 \rho^{-3}}
-\frac{A_{O4}}{\rho\tau^3 }-\frac{A_{O6}}{\tau^3 }
\end{eqnarray}
Extremization Conditions give
\begin{itemize}
 \item $V_{H_{3}}=\frac{13}{2} V_{eff}|_{extr}-2 V_{R_6}+ \frac{1}{2}V_{O6}$
 \item $V_{F_{0}}=\frac{7}{2}V_{eff}|_{extr}+ \frac{1}{2}V_{O6}-V_{R_6}$
 \item $V_{O_{4}}=9 V_{eff}|_{extr}-2 V_{R_6}$
\end{itemize}
The matrix elements are given by 
\begin{eqnarray}
  M_{11} &=& \frac{\partial^2 V_{eff}}{\partial \hat{\rho}^2} = \frac{2}{3}(9 V_{H_{3}} + V_{R_{6}} + 9 V_{F_{0}}-V_{O_{4}}) \nonumber\\
 M_{22} &=& \frac{\partial^2 V_{eff}}{\partial \hat{\tau}^2}= \frac{1}{2}(4 V_{H_{3}} +4 V_{R_{6}} +16 V_{F_{0}} -9 V_{O_{4}}-9 V_{O_{6}}) \nonumber\\
 M_{12} &=& \frac{\partial}{\partial \hat{\tau}}(\frac{\partial V_{eff}}{\partial \hat{\rho}})=\frac{1}{\sqrt{3}}(6 V_{H_{3}}+2 V_{R_{6}} -12 V_{F_{0}}-3 V_{O_{4}}) 
\end{eqnarray}
and 
\begin{eqnarray}
  {\rm Tr}[M]_{extr} &=& \frac{109}{2} V_{eff} +\frac{13}{2} V_{O_{6}}-17 V_{R_{6}}\nonumber\\
  ({\rm Tr}[M]^2-4 {\rm Det}[M])_{extr}&=& \frac{16249}{4} V_{eff}^2 +\frac{169}{4} V_{O_{6}}^2 + \frac{931}{3} V_{R_{6}}^2 +\frac{1657}{2} V_{eff} V_{O_{6}}\nonumber\\
 && -2245 V_{eff} V_{R_{6}}-229 V_{O_{6}} V_{R_{6}}
\end{eqnarray}


\subsection{IIB: $R_6, H_3, F_1, F_3, O_3$}
\begin{eqnarray}
V_{eff} &=& V_{R_{6}} + V_{H_{3}} + V_{F_{1}}+ V_{F_{3}}-V_{O_{3}}\nonumber\\
&=& \frac{A_{R_6}}{\tau^2\rho} + \frac{A_{H_{3}}}{\tau^2 \rho^3} + \frac{A_{F_{1}}}{\tau^4 \rho^{-2}}+ \frac{A_{F_{3}}}{\tau^4 }
-\frac{A_{O3}}{\rho^{3/2} \tau^3 }
\end{eqnarray}
Extremization gives
\begin{itemize}
 \item $V_{H_{3}}=2 V_{eff}|_{extr}-V_{R_6}+ \frac{1}{2}V_{O3}$
 \item $V_{F_{1}}=3 V_{eff}|_{extr}-V_{R_6}$
 \item $V_{F_{3}}=-4 V_{eff}|_{extr}+ V_{R_6}+ \frac{1}{2}V_{O3}$
\end{itemize}
The matrix elements are given by 
\begin{eqnarray}
  M_{11} &=& \frac{\partial^2 V_{eff}}{\partial \hat{\rho}^2} = 6 V_{H_{3}} +\frac{2}{3} V_{R_{6}} + \frac{8}{3} V_{F_{1}}-\frac{3}{2} V_{O_{3}} \nonumber\\
 M_{22} &=& \frac{\partial^2 V_{eff}}{\partial \hat{\tau}^2}= 2 V_{H_{3}} +2 V_{R_{6}} +8 V_{F_{1}}+ 8 V_{F_{3}}-\frac{9}{2} V_{O_{3}} \nonumber\\
 M_{12} &=& \frac{\partial}{\partial \hat{\tau}}(\frac{\partial V_{eff}}{\partial \hat{\rho}})=2\sqrt{3} V_{H_{3}}+\frac{2}{\sqrt{3}} V_{R_{6}} 
 -\frac{8}{\sqrt{3}} V_{F_{1}}-\frac{3\sqrt{3}}{2} V_{O_{3}} 
\end{eqnarray}
with 
\begin{eqnarray}
  {\rm Tr}[M]_{extr} &=& 16 V_{eff} +2 V_{O_{3}}-8 V_{R_{6}}\nonumber\\
  ({\rm Tr}[M]^2-4 {\rm Det}[M])_{extr}&=& 768 V_{eff}^2 + 4 V_{O_{3}}^2 + \frac{256}{3} V_{R_{6}}^2 +96 V_{eff} V_{O_{3}}\nonumber\\
 && -512 V_{eff} V_{R_{6}}-32 V_{O_{3}} V_{R_{6}}
\end{eqnarray}



\subsection{IIB: $ R_6, H_3, F_1, O_3, O_5$}
\begin{eqnarray}
V_{eff} &=& V_{R_{6}} + V_{H_{3}} + V_{F_{1}}- V_{O_{3}}-V_{O5}\nonumber\\
&=& \frac{A_{R_6}}{\tau^2\rho} + \frac{A_{H_{3}}}{\tau^2 \rho^3} + \frac{A_{F_{1}}}{\tau^4 \rho^{-2}}
-\frac{A_{O_{3}}}{\rho^{3/2} \tau^3 }-\frac{A_{O5}}{\sqrt{\rho}\tau^3 }
\end{eqnarray}
Extremization conditions give
\begin{itemize}
 \item $V_{H_{3}}=6 V_{eff}|_{extr}-2 V_{R_6}+ \frac{1}{2}V_{O5}$
 \item $V_{F_{1}}= 3 V_{eff}|_{extr}-V_{R_6}+ \frac{1}{2}V_{O5}$
 \item $V_{O_{3}}=8 V_{eff}|_{extr}-2 V_{R_6}$
\end{itemize}
The matrix elements are given by 
\begin{eqnarray}
  M_{11} &=& \frac{\partial^2 V_{eff}}{\partial \hat{\rho}^2} = 6 V_{H_{3}} +\frac{2}{3} V_{R_{6}} + \frac{8}{3} V_{F_{1}}-\frac{3}{2} V_{O_{3}}
  -\frac{1}{6} V_{O_{5}}\nonumber\\
 M_{22} &=& \frac{\partial^2 V_{eff}}{\partial \hat{\tau}^2}= 2 V_{H_{3}} +2 V_{R_{6}} +8 V_{F_{1}}-\frac{9}{2} V_{O_{3}}
  -\frac{9}{2} V_{O_{5}}\nonumber\\
 M_{12} &=& \frac{\partial}{\partial \hat{\tau}}(\frac{\partial V_{eff}}{\partial \hat{\rho}})=2\sqrt{3} V_{H_{3}}+\frac{2}{\sqrt{3}} V_{R_{6}} 
 -\frac{8}{\sqrt{3}} V_{F_{1}}-\frac{3\sqrt{3}}{2} V_{O_{3}} -\frac{\sqrt{3}}{2} V_{O_{5}}\nonumber
\end{eqnarray}
with
\begin{eqnarray}
  {\rm Tr}[M]_{extr} &=& 32 V_{eff}-12 V_{R_{6}} +\frac{14}{3} V_{O_{5}}\nonumber\\
  ({\rm Tr}[M]^2-4 {\rm Det}[M])_{extr}&=& 1792 V_{eff}^2 +\frac{196}{9} V_{O_{5}}^2 + \frac{496}{3} V_{R_{6}}^2 +\frac{1184}{3} V_{eff} V_{O_{5}}\nonumber\\
 && -1088 V_{eff} V_{R_{6}}-120 V_{O_{5}} V_{R_{6}}
\end{eqnarray}

\subsection{IIB: $ R_6, H_3, F_1, O_3, O_7$}
\begin{eqnarray}
V_{eff} &=& V_{R_{6}} + V_{H_{3}} + V_{F_{1}}- V_{O_{3}}-V_{O7}\nonumber\\
&=& \frac{A_{R_6}}{\tau^2\rho} + \frac{A_{H_{3}}}{\tau^2 \rho^3} + \frac{A_{F_{1}}}{\tau^4 \rho^{-2}}
-\frac{A_{O_{3}}}{\rho^{3/2} \tau^3 }-\frac{A_{O7}}{\rho^{-1/2}\tau^3 }
\end{eqnarray}
Extremization gives
\begin{itemize}
 \item $V_{H_{3}}=6 V_{eff}|_{extr}-2 V_{R_6}+ V_{O7}$
 \item $V_{F_{1}}= 3 V_{eff}|_{extr}-V_{R_6}+ V_{O7}$
 \item $V_{O_{3}}=8 V_{eff}|_{extr}-2 V_{R_6} + V_{O_{7}}$
\end{itemize}
The matrix elements are given by 
\begin{eqnarray}
  M_{11} &=& \frac{\partial^2 V_{eff}}{\partial \hat{\rho}^2} = 6 V_{H_{3}} +\frac{2}{3} V_{R_{6}} + \frac{8}{3} V_{F_{1}}-\frac{3}{2} V_{O_{3}}
  -\frac{1}{6} V_{O_{7}}\nonumber\\
 M_{22} &=& \frac{\partial^2 V_{eff}}{\partial \hat{\tau}^2}= 2 V_{H_{3}} +2 V_{R_{6}} +8 V_{F_{1}}-\frac{9}{2} V_{O_{3}}
  -\frac{9}{2} V_{O_{7}}\nonumber\\
 M_{12} &=& \frac{\partial}{\partial \hat{\tau}}(\frac{\partial V_{eff}}{\partial \hat{\rho}})=2\sqrt{3} V_{H_{3}}+\frac{2}{\sqrt{3}} V_{R_{6}} 
 -\frac{8}{\sqrt{3}} V_{F_{1}}-\frac{3\sqrt{3}}{2} V_{O_{3}} +\frac{\sqrt{3}}{2} V_{O_{7}}\nonumber
\end{eqnarray}
with 
\begin{eqnarray}
  {\rm Tr}[M]_{extr} &=& 32 V_{eff}-12 V_{R_{6}} +8 V_{O_{7}}\nonumber\\
  ({\rm Tr}[M]^2-4 {\rm Det}[M])_{extr}&=& 1792 V_{eff}^2 +\frac{208}{3} V_{O_{7}}^2 + \frac{496}{3} V_{R_{6}}^2 +704 V_{eff} V_{O_{7}}\nonumber\\
 && -1088 V_{eff} V_{R_{6}}-\frac{640}{3} V_{O_{7}} V_{R_{6}}
\end{eqnarray}

\subsection{IIB: $R_6, H_3, F_3, O_3, O_7$}
\begin{eqnarray}
V_{eff} &=& V_{R_{6}} + V_{H_{3}} + V_{F_{3}}- V_{O_{3}}-V_{O7}\nonumber\\
&=& \frac{A_{R_6}}{\tau^2\rho} + \frac{A_{H_{3}}}{\tau^2 \rho^3} + \frac{A_{F_{3}}}{\tau^4 }
-\frac{A_{O_{3}}}{\rho^{3/2} \tau^3 }-\frac{A_{O7}}{\rho^{-1/2}\tau^3 }
\end{eqnarray}
Extremization gives
\begin{itemize}
 \item $V_{F_{3}}=-3 V_{eff}|_{extr}+ V_{R_6}+ V_{H_{3}}$
 \item $V_{O_{3}}= -V_{eff}|_{extr}+ V_{R_6}+ 2 V_{H_{3}}$
 \item $V_{O_{7}}= -3 V_{eff}|_{extr}+ V_{R_6}$
\end{itemize}
The matrix elements are given by 
\begin{eqnarray}
  M_{11} &=& \frac{\partial^2 V_{eff}}{\partial \hat{\rho}^2} = 6 V_{H_{3}} +\frac{2}{3} V_{R_{6}} -\frac{3}{2} V_{O_{3}}
  -\frac{1}{6} V_{O_{7}}\nonumber\\
 M_{22} &=& \frac{\partial^2 V_{eff}}{\partial \hat{\tau}^2}= 2 V_{H_{3}} +2 V_{R_{6}} +8 V_{F_{3}}-\frac{9}{2} V_{O_{3}}
  -\frac{9}{2} V_{O_{7}}\nonumber\\
 M_{12} &=& \frac{\partial}{\partial \hat{\tau}}(\frac{\partial V_{eff}}{\partial \hat{\rho}})=2\sqrt{3} V_{H_{3}}+\frac{2}{\sqrt{3}} V_{R_{6}} 
 -\frac{3\sqrt{3}}{2} V_{O_{3}} +\frac{\sqrt{3}}{2} V_{O_{7}}\nonumber
\end{eqnarray}
with 
\begin{eqnarray}
  {\rm Tr}[M]_{extr} &=& -4 V_{eff} + 4 V_{H_{3}}\nonumber\\
  ({\rm Tr}[M]^2-4 {\rm Det}[M])_{extr}&=& 64 V_{eff}^2 +16 V_{H_{3}}^2 + \frac{16}{3} V_{R_{6}}^2 +32 V_{eff} V_{H_{3}}\nonumber\\
 && -32 V_{eff} V_{R_{6}}
\end{eqnarray}

\end{document}